\documentclass[12pt]{article}

\usepackage{amsmath}
\usepackage{amssymb}
\usepackage{verbatim}

\makeatletter

   \@addtoreset{equation}{section}
\makeatother

\newcommand{\be}{\begin{equation}}
\newcommand{\ee}{\end{equation}}
\newcommand{\ba}{\begin{eqnarray}}
\newcommand{\ea}{\end{eqnarray}}

\begin{document}

\vskip 12mm

\begin{center}

{\Large \bf Higher-Spin Quartic Vertices in $AdS_5$: a Rectangular Limit}
\vskip 10mm
{ \large  Dmitri Polyakov$^{a,b,}$\footnote{email:polyakov@scu.edu.cn;polyakov@sogang.ac.kr}
and  Cong Zhang$^{a}$}

\vskip 8mm
$^{a}$ {\it  Center for Theoretical Physics, College of Physical Science and Technology}\\
{\it  Sichuan University, Chengdu 6100064, China}\\
\vskip 2mm

$^{b}$ {\it Institute of Information Transmission Problems (IITP)}\\
{\it  Bolshoi Karetny per. 19/1, Moscow 127994, Russia}

\end{center}

\vskip 15mm

\begin{abstract}

We determine the form of a significantly large class of gauge-invariant quartic
vertices for symmetric higher-spin fields (in Vasiliev's formalism) in AdS space in the ``rectangular limit''
defined as $s_1+s_2={s_3+s_4+3}$, using the higher-spin 
vertex operators in superstring theory, which construction is based on the generators of the higher-spin symmetry algebra,
enveloping the AdS isometries. In this limit, a particular simplification is that the quartic vertices do not receive corrections from
worldsheet renormalization group flows of the cubic terms, so their structure is fixed   directly by the $4$-point amplitudes
in superstring theory.

\end{abstract}

\vskip 12mm

\setcounter{footnote}{0}

\section{\bf Introduction}

$AdS$ geometry  is a natural background for higher-spin fields to live
since, in the absence of boundary S-matrix, higher-spin gauge theories circumvent
the restrictions of Coleman-Mandula theorem and can be formulated consistently.
Still, describing and classifying higher-spin interactions in $AdS$
is a highly nontrivial problem even at the cubic level and, at levels
higher than cubic, despite some isolated examples of gauge-invariant 
vertices, no classification is known (for an incomplete
list of relevant literature, see e.g.
\cite{fronsdal,  sagnottia, sagnottib, sagnottic,
sagnottid, sorokin, 
mva, mvb, mvc, mvasiliev, deser, siegel, siegelb,
nicolai, damour, brink, boulanger,
labastidaa, labastidab, mvd, mve, mvasiliev, mirian, bekaert, 
perf, pers, taronna,  bekaert, giombif, 
giombis,
sezgin, sundborg, per, zhenya}
Higher-Spin modes in $AdS$ also constitute an important ingredient
of holography and $AdS/CFT$ correspondence, as they are dual to most
of higher-derivative and composite operators on the CFT side
\cite{klebanov, per, giombif, giombis}. In particular,
cubic ingteractions of higher-spin fields in $AdS$  define the structure constants
involving such operators in CFT, while
 quartic interactions in $AdS$ space are holographically related
to conformal blocks on the AdS boundary. These interactions are known
to be genuinely nonlocal and these nonlocalities, in general, cannot
be removed by field redefinitions. The quartic higher-spin interactions 
have a notoriously cumbersome structure and , despite some limited examples
of gauge-invariant $4$-vertices, the classification of quartic interactions
in higher-spin theories largely remains an open problem

Our goal in this paper is to understand
the origin of these nonlocalities from string theory point of view.
String theory has long been understood as a natural framework for the higher-spin fields
since massive vertex operators in string theory have spin values of the order $s\sim{m^2}$
on the leading Regge trajectory. These operators formally become massless in the tensionless limit,
however, there is no easy way to relate this limit to string perturmation theory
and the low-energy effective action which could have shed light on higher-spin interactions.
In addition, the higher-spin vertex operators in the spectrum of conventional string theory
do not possess sufficient gauge symmetries , necessary to eliminate 
So despite some manifest higher-spin structures appearing in standard string theory, there is no obvious
way to relate interacting higher-spin theories with full gauge symmetries to conventional string dynamics.
One reason for this is that conventional string theory lacks the authentic space-time symmetries 
present in higher-spin theories: that is , if a higher-spin theory lives in a space-time 
with isometry algebra $G$, then, in  Vasiliev's frame-like description,
 one expects such a theory to be a realization of infinite-dimensional symmetry algebra which is the
enveloping of $G$ (higher-spin algebra). Conventional string theory clearly has no such isometries,
with its space-time symmetry algebra limited to Poincare or $AdS$. This is related to the fact that
conventional string theory can be thought of as a special phase of higher-spin theories,
with the higher-spin symmetries broken to finite-dimensional space-time isometries.
For example, in bosonic string theory (which we shall consider for a moment for simplicity)
the Poincare isometries are realized by the translation and rotation operators  given by:

\begin{eqnarray}
T^m=\oint{{dz}\over{2i\pi}}\partial{X^m}(z)
\nonumber \\
T^{mn}=\oint{{dz}\over{2i\pi}}X^{{\lbrack}m}\partial{X^{n\rbrack}}(z)
\end{eqnarray}
with $X^{m}; m=0,...D-1$ being coordinates in $D$-dimensional target space and $(z,{\bar{z}})$
parametrizing the coordinates in  conformal or superconformal gauges.
In RNS superstring theory in flat space, the same Poincare algebra can be realized
provided that the roration/boost generators are modified according to
\begin{equation}
T^{mn}\rightarrow{T^{mn}}+\oint{{dz}\over{2i\pi}}\psi^m\psi^n
\end{equation}
where $\psi^m$ are worldsheet RNS fermions.

There are a few useful observations to be made here.
First of all, the space-time isometry generatiors are the physical objects,
i.e. they are the worldsheet integrals of dimension 1 primary fields and are in the BRST cohomology.
This is the common property of the space-time symmetry generators in string theory.
Second, they are structurally related to massless physical excitations  in open string or superstring theory;
multiplied by exponential fields  $\sim{e^{ipX}}$ and the appropriate
polarization vector, they produce vertex operators for spin $1$ vertex operators in the string spectrum at the 
space-time momentum $p$.
Thus the space-time isometries are closely related to the dynamics of massless particles emitted by open strings;
in the simplest case those are spin one masless gauge bosons (photons).
At the same time , the massiveness of higher-spin operators in  string theory is a hint that string theory
is actually a theory with  spontaneously  broken higher-spin symmetries.
These symmetries can be of course restored by hands, that is, by introducing Stuckelberg fields.
But this by itself does not solve the problem of the negative norm modes, since Stuckelberg variables with themselves
bring extra degrees of freedom that gauge symmetries aim to eliminate.
The question is then - can one modify or ``enlarge'' string theory so that the enlarged theory contains the authentic 
space-time symmetries that are broken in the standard formulation? The answer is positive, 
and RNS superstring theory appears to be
a particularly efficient framework to elaborate.
The appearance of higher-spin algebra in superstring theory is closely related to 
the operators in higher  ghost cohomologies, related to global singularities in  the supermoduli space.
These operators , up to BRST-exact terms, commute trivially with conventional Poincare generators,
that by definitions are the elements of the zero cohomology $H_0$.
The operators, realizing hidden $AdS$ space-time isometries in $RNS$ superstring theory, are the elements of
the first nontrivial cohomology $H_1\sim{H_{-3}}$, given by:

\begin{eqnarray}
L_{+}^m(w)=K\circ{P^m}
\nonumber \\
=\oint{{dz}\over{2i\pi}}(z-w)^2
\lbrace{1\over2}B^{(2)}_{2\phi-2\chi-\sigma}{e^\phi}F^m_{5\over2}
-12\partial{c}ce^{2\chi-\phi}
F^m_{5\over2}\nonumber \\
+ce^\chi\lbrack
-{2\over3}\partial^3\psi^m\lambda+{4\over3}\partial^3\varphi\partial{X^m}
+2\partial^2\psi^m\partial\lambda
\nonumber \\
+B^{(1)}_{\phi-\chi}(-2\partial\varphi\partial^2{X^m}
+4\partial^2\varphi\partial{X^m}-2\partial^2\psi^m\lambda
+4\partial\psi^m\partial\lambda)
\nonumber \\
+B^{(2)}_{\phi-\chi}(2\partial\varphi\partial{X^m}+2\psi^m\partial\lambda-
2\partial\psi^m\lambda-q\partial^2{X^m})
\nonumber \\
+B^{(3)}_{\phi-\chi}(-{2\over3}\psi^m\lambda+{{4q}\over3}\partial{X^m})\rbrack
\rbrace
\nonumber \\
=-4{\lbrace}Q,
\oint{{dz}\over{2i\pi}}(z-w)^2
ce^{2\chi-\phi}F^m_{5\over2}(z)\rbrace
\nonumber \\
P^m=\oint{{dz}\over{2i\pi}}e^{\phi}F^m(z)
\nonumber \\
L_{-}^m=\oint{{dz}\over{2i\pi}}e^{-3\phi}F^m(z)
\end{eqnarray}

where $L_{+}^m$ and $L_{-}^m$ are the isomorphic AdS transvection generators in isomorphic positive and negative $H_1$ and $H_{-3}$ 
cohomologies respectively,

\begin{align}
F^m_{5\over2}=\lambda\partial^2{X_m}-2\partial\lambda\partial{X_m}
\end{align}
is dimension ${{5}\over2}$ matter primary, $K\circ...$ is the homotopy transformation
to ensure the BRST-invariance

and the two-form ``rotation'' operator is

\begin{align}
L_{mn}=K\circ{P_{mn}}
\nonumber \\
=\oint{{dz}\over{2i\pi}}
{\lbrack}\psi_m\psi_n+2ce^{\chi-\phi}\psi_{\lbrack{m}}\partial{X_{n\rbrack}}
-4\partial{c}ce^{2\phi-2\chi}\rbrack\nonumber \\
=-4\lbrace{Q},\xi{\Gamma^{-1}}\psi_m\psi_n\rbrace
\nonumber \\
P_{mn}=\oint{{dz}\over{2i\pi}}\psi_m\psi_n
\end{align}

Altogether, $L^m_{\pm},L^{mn}$ operators realize the AdS isometry algebra, e.g. it is straightforward to show that:
\begin{eqnarray}
:\Gamma^2\lbrack{L_{+}^m,L_{-}^n}\rbrack:=-L^{mn}
\end{eqnarray}
where $\Gamma$ is picture-changing operator, and with all other commutations 
identical to AdS/Poincare algebra, up to BRST-exact terms \cite{selfframes}.

Based on these symmetry  generators, one can construct open string physical vertex operators in BRST cohomology which beta-functions
describe photons propagating in AdS background and closed string vertex operators describing the spin 2 massless field in the frame approach
of MMSW gravity \cite{selfframes}
Higher order ghost cohomologies $H_n\sim{H_{-n-2}};n>1$ provide natural infinite-dimensional enveloping of this algebra,
that is, the higher-0spin algebra in $AdS$.
Namely, introduce a higher-spin extension of Cartan 1-form:
\begin{eqnarray}
\Omega^{(1)}=\Omega_m{dx^m}
\nonumber \\
\Omega_m=L_a{e^a_m}(x)+L_{ab}\omega_m^{ab}(x)+\sum_{s=2}^\infty\sum_{t=0}^{s-1}H^{a_1...a_{s-1}|b_1...b_t}_m(x)L_{a_1...a_{s-1}|b_1...b_t}
\end{eqnarray}
where $m$ indices parametrize curved space-time, $a,b$ indices parametrize the tangent space,
$e$ and $\omega$ are spin $2$  vielbein and spin connection fields, while $H$ are the two-row higher-spin fields
in the frame-like desctiption.
The generators $L_a$ and $L_{ab}$ parametrize the isometry algebra of the space-time (e.g. $AdS$) while
$L_{a_1...a_{s-1}|b_1...b_t}$-operators form the higher-spin enveloping of these isometries and the higher-spin algebra, which 
maximal finite subalgebra is the isometry of the underlying space-time.
There is a natural realization of these operators and, accordingly, of the higher-spin algebra
 in RNS superstring theoty , with the spin values corresponding to the ghost cohomology ranks.
The expression for the higher-spin algebra generators become particularly
simple for $t=s-3$.
In the negative cohomology representation, the generators are given by
\begin{eqnarray}
T_{a_1...a_{a_{s-1}}|m;b_1...b_{s-3}}=\oint{{dz}\over{2i\pi}}e^{-s\phi}\partial{X^{a_1}}...\partial{X^{a_{s-1}}}\psi_{m}
\partial\psi_{b_1}...\partial^{s-3}\psi_{b_{s-3}}
\end{eqnarray}
and
\begin{eqnarray}
T_{a_1...a_{a_{s-1}}|m;b_1...b_{s-3}}=K\circ\oint{{dz}\over{2i\pi}}e^{(s-2)
\phi}\partial{X^{a_1}}...\partial{X^{a_{s-1}}}\psi_{m}
\partial\psi_{b_1}...\partial^{s-3}\psi_{b_{s-3}}
\end{eqnarray}
in the positive cohomology representation.
The corresponding higher-spin vertex operators for the space-time two-row
higher-spin fields in the frame-like formalism are given by
\begin{eqnarray}
V_{s|s-3}=H_{a_1...a_{s-1}|m;b_1...b_{t-3}(p)}\oint{{dz}\over{2i\pi}}e^{-s\phi}\partial{X^{a_1}}...\partial{X^{a_{s-1}}}\psi_{m}
\partial\psi_{b_1}...\partial^{s-3}\psi_{b_{s-3}}e^{ipX}(z)
\nonumber \\
\equiv\Omega_{s|s-3}W_{s|s-3}^{(-)}
\end{eqnarray}
in the negative $H_{-s}$-cohomology
and
\begin{eqnarray}
V_{s|s-3}^{(+)}=K\circ{H_{a_1...a_{s-1}|m;b_1...b_{t-3}}}\oint{{dz}\over{2i\pi}}e^{(s-2)                                                      
\phi}\partial{X^{a_1}}...\partial{X^{a_{s-1}}}\psi_{m}
\partial\psi_{b_1}...\partial^{s-3}\psi_{b_{s-3}}e^{ipX}(z)
\nonumber \\
\equiv\Omega_{s|s-3}(p)W_{s|s-3}^{(+)}
\end{eqnarray}
in the dual  positive $H_{s-2}$-cohomology where,
in our notations, $W$ are the worldsheet operators 
multiplied by frame-like space-time fields $\Omega_{s|t}(p)$, making the complete vertex operators $V_{s|t}$ 
The explicit expressions for the operators/ higher spin algebra generators $V_{s|t}$ for $t\neq{s-3}$
are more complicated and can be cast as solutions of the operator equations:
\begin{eqnarray}
::\Gamma^{\pm(s-3-t)}\Omega_{s|t}(p):W^{(\pm)}_{s|t}:= \Omega_{s|s-3}(p)W^{(\pm)}_{s|s-3}
\end{eqnarray}
where $:\Gamma^n:$ are the powers of the picture-changing operator.
The operator equations (1.12) particularly produce the generalized zero-torsion
constraints on the extra fields $\Omega_{s|t}$, relating them to the dynamical field $\Omega_{s|0}$ in the
frame-like formalism. In general, these equations are hard to solve manifestly because of complexity
of both $:\Gamma^n:$'s structure and the operatror products involved.
One particular example of such solution was demonstrated in 
\cite{selfframes} for the case $t=s-4$, and even in that
simplest case the solution found was quite tedious. As the difference between $s-3$ and $t$ increases, so does the 
complexity of the operator equations (1.12). Nevertheless, these equations do provide some useful information.
Namely, it is straightforward to show that the leading order 
(linearized) Weyl invariance constraints on $\Omega_{s|s-3}$ space-time
fields are  given by
\begin{eqnarray}
0=\beta_m^{a_1...a_{s-1}}=
-p^2\Omega_m^{a_1...a_{s-1}}(p)
+
{\Sigma_1}(a_1|a_2,...a_{s-1})p_tp^{a_1}\Omega_m^{a_2...a_{s-1}t}
\nonumber \\
-{1\over2}\Sigma_2(a_{s-2},a_{s-1}|a_1,...,a_{s-3})p^{a_{s-1}}p^{a_{s-2}}
(\Omega_m^\prime)^{a_1...a_{s-3}}
-4(s-1)\Omega_m^{a_1...a_{s-1}}
\end{eqnarray}
The appearance of the mass-like terms is  related to the non-trivial ghost dependence
of the operators (1.10)-(1.12) and points to the emergence
of the curved $AdS$ geometry, related to the hidden space-time symmetries realized
by higher ghost cohomologies. 
In case when the  higher-spin modes are propagating in four-dimensional subspace
(which would correspon d to the transverse $AdS$ directions), the
 vanishing of the $\beta$-function gives, in the leading order,
the low-energy  effective equations of motion on $\Omega$
given by
\begin{equation}
{\hat{F}}_{AdS}\Omega=0
\end{equation}
${\hat{F}}_{AdS}$ is the Fronsdal's operator in $AdS_{d+1}$ space
(exact for $d=4$ and with some modifications in other dimensions),
given by in  the position space:

\begin{eqnarray}
({\hat{F}}_{AdS}\Omega)^{a_1...a_s}
=\nabla_A\nabla^A\Omega^{a_1...a_s}-\Sigma_1
(a_1|a_2...a_s)\nabla_t\nabla^{(a_1}
\Omega^{a_2...a_st)}
\nonumber \\
+{1\over2}\Sigma_2(a_1,a_2|a_3,...,a_s)
\nabla^{a_1}\nabla^{a_2}(\Omega^\prime)^{a_3...a_s}
-m_\Omega^2\Omega^{a_1...a_s}+2\Sigma_2\Lambda{g^{a_1a_2}}(\Omega^\prime)^{a_3...a_s}
\nonumber \\
m^2_\Omega=-\Lambda(s-1)(s+d-3)
\end{eqnarray}
where $A=(a,\alpha)$ is the $AdS_{d+1}$ space-time index
(with the latin indices being along the boundary and $\alpha$
being the radial direction), and $\Sigma_1,\Sigma_2$ are the Fronsdal's
symmetrization generators.
In what follows, we shall limit ourselves to the $d=4$ case, in order 
to simplify things.

\section{\bf  Quartic Interactions and Nonlocalities}

In string theory,
the structure of the higher-spin quartic interaction is related to the
four-point worldsheet correlators. Typically,  
the four point correlators of ``standard'' vertex operators  (elements
of $H_0$) 
lead to  Veneziano amplitudes, bearing no trace of nonlocality
and leading to perfectly local quartic terms in the low-energy effective 
action.
In case of the amplitudes involving the operatorsof $H_n\sim{H_{n-2}}$
for higher-spin fields, the situation changes radically 
because of the different $b-c$ ghost content of these operators.
That is,consider the standard Veneziano amplitude. Its well-known structure
results from the four-point function involving one 
integrated and three unintegrated vertices, that ensure the cancellation
of the $b-c$ ghost number anomaly due to the background charge of the $b-c$
 system. Three $c$-ghost insertions on the sphere lead to the standard
$SL(2,R)$ volume factor for open strings and $SL(2,C)$ for closed strings.
The single worldsheet integration then leads to the amplitude structure
$\sim{{\Gamma\Gamma}\over{\Gamma}}$ for open strings
and $\sim{{\Gamma\Gamma\Gamma}\over{\Gamma\Gamma\Gamma}}$ for closed strings,
where $\Gamma$ are the gamma-functions in Mandelstam variables.
The simple poles in the amplitudes occur at non-positive integer values
of the Mandelstam variables; in particular, residues at massless
poles  determine the quartic interactions of space-time field
in the low-energy effective action (in the leading order of $\alpha^\prime$).
Things change significantly with the higher-spin operators entering the game.
Typically, a four-point amplitude involving the spin $n+2$ operators
of $H_n\sim{H_{-n-2}}(n>0)$ must contain at least two vertex operators
at positive picture representation, in order to cancel the 
background charge of the $\beta-\gamma$ system, equal to $2$.
As we have seen above, such operators do not admit a representation
at unintegrated $b-c$ picture, so the $4$-point amplitudes of the
 higher-spin operators involve at least $2$ worldsheet integrations.
Because of that, the resulting expressions for the amplitudes
develop the ``anomalous'' factors (in addition to the standard Veneziano
structure) leading to the appearance of the nonlocalities in the quartic
interactions of the higher-spin fields.
Below we shall consider a few examples of how this scenario for the
higher-spin nonlocalities unfolds in practice.
Consider the 4-point amplitude, describing the interaction of
massless higher-spin fields with the spin values $s_1$,$s_2$, $s_3$ and $s_4$.
The structure of this amplitude becomes relatively simple if the spin values are
subject to the constraint:
\begin{align}
s_1+s_2=s_3+s_4+3
\end{align}
If this constraint is satisfied, the operators (1.10), 
with the linearized on-shell constraints on the space-time fields
describing the propagation of massless 
frame-like higher-spin modes in AdS (along the AdS boundary),
can be taken at their canonical pictures. Otherwise, 
 the calculation of the amplitude would
require the insertions of the picture-changing operators, related to the generalized
zero-torsion constraints for frame-like fields, making the whole computation quite messy.
Nevertheless, the special case (2.1) is already general enough 
to grasp the  architectural basics of the higher-spin quartic interactions in the AdS
space.
With the constraints (2.1) satisfied, 
it is natural to choose the spins $s_1$ and $s_2$ operators 
to be in positive cohomologies and those of spins $s_3$ 
and $s_4$ $b-c$ local and in the negative cohomologies,
so that the open string amplitude for the quartic has the form:

\begin{align}
A(s_1...s_4|p_1...p_4)
\nonumber \\
=<K\circ{\int_0^1{d\xi_1}{e^{(s_1-2)\phi}}}\prod_{i,j=1}^{s_1-2}\partial^{i-1}\psi_{\mu_i}\partial{X}_{\alpha_j}e^{ip_1X}(\xi_1)
\nonumber \\
K\circ{\int_0^{\xi_1}{d\xi_2}{e^{(s_2-2)\phi}}}\prod_{i,j=1}^{s_2-2}\partial^{i-1}\psi_{\nu_i}\partial{X}_{\beta_j}e^{ip_2X}(\xi_2)
\nonumber \\
c{{e^{-s_3\phi}}}\prod_{i,j=1}^{s_3-2}\partial^{i-1}\psi_{\rho_i}\partial{X}_{\gamma_j}e^{ip_3X}(\xi_3=0)
\nonumber \\
c{{e^{-s_4\phi}}}\prod_{i,j=1}^{s_4-2}\partial^{i-1}\psi_{\sigma_i}\partial{X}_{\delta_j}e^{ip_4X}(w\rightarrow\infty)
>
\end{align}
where we have partially fixed  the $SL(2,R)$ symmetry , choosing the negative cohomology operators
at $0$ and $\infty$.
Now it is clear that, according to the ghost number selection rules, up to the interchange of $\xi_1\leftrightarrow\xi_2$
the $s_1$ and $s_2$ operators each of two $K$-transformation's  only contributions to the correlator are:

1) the one proportional to $\sim{e^{\chi+(s_1-3)\phi}}$ ghost factor in $K\circ{V_{s_1}}$.

2)the one proportional  to $\sim{e^{(s_1-2)\phi}}$ ghost factor in $K\circ{V_{s_2}}$.

The straightforward evaluation of these factors in the $K$-transformations,
using the symmetry in $\alpha$ and $\mu$ indices, 
gives:
\begin{align}
K\circ{V_s}(p)=V_s^{(1)}(p)+V_s^{(2)}(p)
V_s^{(1)}(\zeta)
\nonumber \\
=-{1\over2}\Omega^{\alpha_({s-1})|\mu_{(s-3)}}
(p)
\int{dz}(\zeta-z)^{2s-4}:e^{ipX}
\nonumber \\
\times
\lbrace
\sum_{q=1}^{s-2}
(-1)^q(q-1)!\lbrack{T^\perp{\psi}}^{\mu_q}_{\mu(s-2)}\rbrack
\nonumber \\
\times
({\lbrack}T^{||}X{\rbrack}_{\alpha(s-1)}^{\alpha_{s-1}}
\eta^{\mu_q\alpha_{s-1}}
(1-s)B_{\phi-\chi}^{(s+q+2)}
\nonumber \\
+
{\lbrack}T^{||}X{\rbrack}_{s-1}
(-ip)^{\mu_q}
B_{\phi-\chi}^{(s+q+1)})
\nonumber \\
+
\sum_{q=1}^{s-2}\sum_{r=0}^{s+q}
(-1)^q(q-1)!\lbrack{T^\perp{\psi}}^{\mu_q}_{\mu(s-2)}\rbrack
\nonumber \\
{\times}({\lbrack}T^{||}X{\rbrack}_{s-1}
B_{\phi-\chi}^{(s+q-r)}{{\partial^{r+1}X_{\mu_q}}\over{r!}})
\nonumber \\
-
\sum_{r=0}^{s+2}
(-1)^q(q-1)!\lbrack{T^\perp{\psi}}_{\mu(s-2)}\rbrack
\nonumber \\
\times
({\lbrack}T^{||}X{\rbrack}^{\alpha_{s-1}}_{\alpha(s-1)}\rbrack
B_{\phi-\chi}^{(s+2-r)}{{\partial^{r}\psi_{\alpha_{s-1}}}\over{r!}})
\nonumber \\
-
\sum_{r=0}^{s+2}
(-1)^q(q-1)!\lbrack{T^\perp{\psi}}_{\mu(s-2)}\rbrack
\nonumber \\
\times
({\lbrack}T^{||}X{\rbrack}_{\alpha(s-1)}\rbrack
B_{\phi-\chi}^{(s+1-r)}{{p^\alpha\partial^{r}\psi_\alpha}\over{r!}}
\rbrace:(z))
\end{align}
\begin{align}
V_s^{(2)}(\zeta)=-{1\over4}\Omega^{\alpha_({s-1})|\mu_{(s-3)}}
(p)
\nonumber \\
\times
\int{dz}(\zeta-z)^{2s-4}:e^{ipX}
B^{(2s-4)}_{2\phi-2\chi-\sigma}
\lbrack{T^\perp{\psi}}_{\mu(s-2)}\rbrack
\nonumber \\
\times
{\lbrack}T^{||}X{\rbrack}_{\alpha(s-1)}:(z)\rbrack
\end{align}
where we have adopted the following notations:
\begin{align}
{\lbrack}T^{||}X{\rbrack}_{\alpha(s)}=:\prod_{i=1}^s\partial{X_{\alpha_i}}:
\nonumber \\
{\lbrack}T^{||}X{\rbrack}_{\alpha(s)}^{\alpha_q}=:\prod_{i=1;i\neq{q}}^s\partial{X_{\alpha_i}}:
\nonumber \\
\lbrack{T^\perp{\psi}}_{\mu(s)}\rbrack=:\prod_{j=1}^{s}\partial^{j-1}\psi_{\mu_j}:
\nonumber \\
\lbrack{T^\perp{\psi}}_{\mu(s)}^{\mu_q}\rbrack=:\prod_{j=1;j\neq{q}}^{s}\partial^{j-1}\psi_{\mu_j}:
\nonumber \\
\Omega^{\alpha_(s)|\mu_{(t)}}(p)=\Omega^{\alpha_1...\alpha_s|\mu_1...\mu_t}(p)
\end{align}
Given the ghost number selection rules, the terms contributing to
the $4$-point correlator are given by:
$$<V_{s_1}^{(1)}(\zeta_1)V_{s_2}^{(2)}(\zeta_2)V_{s_3}(0)V_{s_2}(w\rightarrow\infty)>+
\xi_1\rightarrow\xi_2;s_1\rightarrow{s_2}$$

The overall correlator stems from the contributions of $\psi$-dependent , $X$-dependent and ghost-dependent
ingredients of the higher-spin vertex operators.
We start from the $\psi$-part.
It is given by the two patterns; the first one is

\begin{align}
<T^{\perp}\psi_{\mu(s_1-2)}^{\mu_q}(\zeta_1)T^{\perp}\psi_{\nu(s_2-2)}(\zeta_2)
T^{\perp}\psi_{\rho(s_3-2)}(0)
\nonumber \\
T^{\perp}\psi_{\sigma(s_4-2)}(w\rightarrow\infty)>
=\sum_{s_1-3,s_2-2,s_3-2,s_4-2|\lbrace\lambda_{ij}\rbrace}
\nonumber \\
\sum_{{{(s_1-2)(s_1-3)}\over2}-q;{{(s_2-2)(s_2-3)}\over2};
{{(s_3-2)(s_2-3)}\over2};{{(s_4-2)(s_4-3)}\over2}|\lbrace|\Delta_{ij}\rbrace}
\nonumber \\
\Xi^{(1-\psi)}_{q;\mu_1...\mu_{s_1-2}\nu_1...\nu_{s_2-2}\rho_1...\rho_{s_3-2}
\sigma_1...\sigma_{s_4-2}}(\lbrace{\lambda\rbrace,\lbrace{\Delta}\rbrace})
\nonumber \\
\times(\zeta_1-\zeta_2)^{-(\Delta_{12}+\delta_{21}+\lambda_{12})}
\zeta_1^{-(\Delta_{13}+\Delta_{31}+\lambda_{13})}
\nonumber \\
\times
\zeta_2^{-(\Delta_{23}+\Delta_{32}+\lambda_{23})}
w^{{{-(s_4-2)(s_4-1)}\over2}-\Delta_{41}-\Delta_{42}-\Delta_{43}}
\nonumber \\
\equiv\sum_{s_1-3,s_2-2,s_3-2,s_4-2|\lbrace\lambda_{ij}\rbrace}
\sum^{(s_1-3;s_2-2;s_3-2;s_4-2)}_{ord.\lbrace{i,j,k}\rbrace,
\lbrace{l,m,n}\rbrace,\lbrace{p,q,r}\rbrace,\lbrace{s,t,u}\rbrace}
\nonumber \\
\sum_{{{(s_1-2)(s_1-3)}\over2}-q;
{{(s_2-2)(s_2-3)}\over2};{{(s_3-2)(s_2-3)}\over2};
{{(s_4-2)(s_4-3)}\over2}|\lbrace|\Delta_{ij}\rbrace}
\nonumber \\
(\zeta_1-\zeta_2)^{\Delta_{12}+\delta_{21}+\lambda_{12}}
\zeta_1^{\Delta_{13}+\Delta_{31}+\lambda_{13}}
\nonumber \\
\times
\zeta_2^{\Delta_{23}+\Delta_{32}+\lambda_{23}}
w^{{{(s_4-2)(s_4-1)}\over2}+\Delta_{41}+\Delta_{42}+\Delta_{43}}
\nonumber \\
\times
\eta_{\mu_{i_1}\nu_{l_1}}...\eta_{\mu_{i_{\lambda_{12}}}\nu_{l_{\lambda_{12}}}}
\eta_{\mu_{j_1}\rho_{p_1}}...\eta_{\mu_{i_{\lambda_{13}}}\rho_{p_{\lambda_{13}}}}
\nonumber \\
\times
\eta_{\mu_{k_1}\sigma_{s_1}}...\eta_{\mu_{k_{\lambda_{14}}}\sigma_{s_{\lambda_{14}}}}
\eta_{\nu_{m_1}\rho_{r_1}}...\eta_{\nu_{m_{\lambda_{23}}}}
\rho_{r_{\lambda_{23}}}
\nonumber \\
\eta_{\nu_{n_1}\sigma_{t_1}}...\eta_{\nu_{n_{\lambda_{24}}}
\rho_{t_{\lambda_{24}}}}
\eta_{\rho_{r_1}\sigma_{u_1}}...\eta_{\rho_{r_{\lambda_{34}}}}
\sigma_{u_{\lambda_{34}}}
\nonumber \\
(i_1+l_1-2)!...(i_{\lambda_{12}}+l_{\lambda_{12}}-2)!
\nonumber \\
(j_1+p_1-2)!...(j_{\lambda_{13}}+p_{\lambda_{13}}-2)!
\nonumber \\
(k_1+s_1-2)!...(k_{\lambda_{14}}+s_{\lambda_{14}}-2)!
\nonumber \\
(m_1+q_1-2)!...(m_{\lambda_{23}}+q_{\lambda_{23}}-2)!
\nonumber \\
(n_1+t_1-2)!...(n_{\lambda_{24}}+t_{\lambda_{24}}-2)!
\nonumber \\
(r_1+u_1-2)!...(r_{\lambda_{34}}+u_{\lambda_{34}}-2)!
\nonumber \\
\times
(-1)^{(s_1+\lambda_{12})s_2+s_3(s_4+\lambda_{34})+\lambda_{12}+\lambda_{34}+\lambda_{13}\lambda_{24}
+\Delta_{23}+\Delta_{24}+\Delta_{34}}
\nonumber \\
(-1)^{\pi^{(s_1-3)}_q(i_1...i_{\lambda_{12}};j_1...j_{\lambda_{13}};k_1...k_{\lambda_{14}})}
\nonumber \\
(-1)^{\pi^{(s_2-2)}({l_1...l_{\lambda_{12}};m_1...m_{\lambda_{23}};n_1...n_{\lambda_{24}}})}
\nonumber \\
(-1)^{\pi^{(s_3-2)}({p_1...p_{\lambda_{13}};q_1...q_{\lambda_{23}};r_1...r_{\lambda_{34}}})}
\nonumber \\
(-1)^{\pi^{(s_4-2)}({s_1...s_{\lambda_{14}};t_1...t_{\lambda_{24}};u_1...u_{\lambda_{34}}})}
\end{align}
where, by definition, the
$\Xi^{(1-\psi)}_{\mu_1...\mu_{s_1-2}\nu_1...\nu_{s_2-2}
\rho_1...\rho_{s_3-2}\sigma_1...\sigma_{s_4-2}}
(\lbrace{\lambda\rbrace,\lbrace{\Delta}\rbrace})$-factor, stemming from the summations 
over the orderings, which structure is  explained below,
 is introduced in the equation above
to abbreviate the notations.
The sums in this formula are taken over the  (non-ordered) partitions and over the orderings, 
with the notations (both in (4.120)
and in the equations below) defined as follows:

1)$\sum_{n_1...n_p|\lbrace\lambda_{ij}\rbrace}$
denotes the sum over all the non-ordered
partitions of $n_j>0;j=1,...p$ in non-negative $\lambda_{ij}=\lambda_{ji}$:
\begin{equation}
n_j=\sum_{k=1;k\neq{j}}^p\lambda_{jk}
\end{equation}
Similarly,
$\sum_{n_1...n_p|\lbrace\Delta_{ij}\rbrace}$
stands for the sum over all the non-ordered partitions of $n_j>0;j=1,...p$
in non-negative  $\Delta_{ij}$ (in general, $\Delta_{ij}\neq{\Delta_{ji}}$)

\begin{equation}
n_j=\sum_{k=1;k\neq{j}}^p\Delta_{jk}
\end{equation}

2)$\sum^{(n_1;...n_p)}_{ord.{\lbrace{k^{(1)}}\rbrace}...\lbrace{k^{(p)}}\rbrace}$
for each $n^{(j)};j=1,...p$
stands for the total $p$ summations over all the possible orderings of $n_j$ natural numbers 
from 1 to $n_j;1\leq{j}\leq{p}$ : ${\lbrack{k^{(j)}_1}...k^{(j)}_{n_j}\rbrack}$
($1\leq{k_q^{(j)}}\leq{n_j}; k_{q_1}^{(j)}\neq{k_{q_2}^{(j)}}$).

3) Similarly, for each 
$\sum^{(n_1;...n_p)}_{ord.{\lbrace{k^{(1)}}\rbrace}_{q_1}...\lbrace{k^{(p)}_{q_j}}\rbrace}$
stands for the total number of $p$ summations over the possible  orderings of $n_j-1$ natural numbers
from $1$ to $n_j$ with $q_j$ omitted.

4)$\pi^{(n)}(i_1,...i_n)$ stands for the number of nearest-neighbour permutations
of a pair of numbers it takes to create the ordering $i_1...i_n$ of n numbers
from $1$ to $n$ from the ordering $1,2,...n$;
$\pi_q^{(n)}(i_1,...i_{n-1})$ stands for the number   of nearest-neighbour permutations
of a pair of numbers it takes to creatye the ordering $i_1...i_{n-1}$ of $n-1$ numbers
from $1$ to $n$ with $q$ omitted from the ordering $1,2,...,q-1,q+1,...n$.
The $\lbrace{i,j,k,l,m,n,p,q,r,s,t,u}\rbrace$ elements of the orderings
in the formula
 satisfy the constraints:

\begin{align}
i_1+...+i_{\lambda_{12}}=\delta_{12}
\nonumber \\
j_1+...+j_{\lambda_{13}}=\delta_{13}
\nonumber \\
k_1+...+k_{\lambda_{14}}=\delta_{14}
\nonumber \\
l_1+...+l_{\lambda_{12}}=\delta_{21}
\nonumber \\
m_1+...+m_{\lambda_{23}}=\delta_{23}
\nonumber \\
n_1+...+j_{\lambda_{24}}=\delta_{24}
\nonumber \\
p_1+...+p_{\lambda_{13}}=\delta_{31}
\nonumber \\
q_1+...+q_{\lambda_{23}}=\delta_{32}
\nonumber \\
r_1+...+r_{\lambda_{34}}=\delta_{34}
\nonumber \\
s_1+...+s_{\lambda_{14}}=\delta_{41}
\nonumber \\
t_1+...+t_{\lambda_{24}}=\delta_{42}
\nonumber \\
u_1+...+u_{\lambda_{34}}=\delta_{43}
\nonumber \\
\sum_{j}\delta_{1j}=(s_1-2)(s_1-3)-q
\nonumber \\
\sum_{j}\delta_{ij}=(s_i-2)(s_i-3)
\end{align}

The partitions described above have a simple meaning in terms of the contractions between the $\psi$-fields
contributing to the correlator (2.2).  For each term, $\lambda_{ij}=\lambda_{ji}$ is the number of contractions
between the higher-spin operators $V_{s_i}$ and $V_{s_j}$ ($i,j=1,...,4;i\neq{j}$. Obviously, for each $i$ 
fixed the sum of $\lambda_{ij}$ over $j$ gives the total number of $\psi$-fields in the $V_{s_i}$ vertex operator
(equal to $s_i-2$ for the operators in negative cohomologies and $s_i-3$ or $s_i-1$ for the operators
in positive cohomologies (upon the $K$-homotopy transformation).
Next, for each term, $h_{ij}=\Delta_{ij}+{1\over2}\lambda_{ij}$ is the conformal dimension
that $V_{s_i}$ operator contributes to contractions with $V_{s_j}$,
with $\Delta_{ij}$ stemming from the contributions from the derivatives and ${1\over2}\lambda_{ij}$
from the total number $\lambda_{ij}$ of the $\psi$-fields themselves.
The sum  of $h_{ij}$ over $j$ obviously equals to the total conformal dimension of the $\psi$-fields
and their derivatives in $V_{s_i}$ and is equal to ${{(s_j-2)^2}\over2}$
Finally, the $(-1)^{...}$-factors , including those of $(-1)^{\pi(...)}$ stem from permutations
of the $\psi$-fields participating in the contractions, as these fields have odd statistics.
The second $\psi$-pattern, contributing to the overall correlator of the higher-spin operators i
given by :
\begin{align}
A_\psi^{(2)}=
<:\partial^r\psi_{\alpha_1}{\lbrace}T^{\perp}\psi_{\mu(s_1-2)}\rbrace:(\zeta_1)
{\lbrace}T^{\perp}\psi_{\nu(s_2-2)}\rbrace(\zeta_2)
\nonumber \\
{\lbrace}T^{\perp}\psi_{\rho(s_3-2)}\rbrace(0)
{\lbrace}T^{\perp}\psi_{\sigma(s_4-2)}\rbrace(w\rightarrow\infty)>
\nonumber \\
=
\sum_{{\lbrace}s_1-1,s_2-2,s_3-2,s_4-2|\lbrace\lambda_{ij}\rbrace\rbrace}
\nonumber \\
\sum_{\lbrace{{(s_1-2)(s_1-3)}\over2}+r;{{(s_2-2)(s_2-3)}\over2};
{{(s_3-2)(s_3-3)}\over2};{{(s_4-2)(s_4-3)}\over2}|\lbrace\Delta_{ij}
\rbrace\rbrace}
\nonumber \\
\lbrace
(\zeta_1-\zeta_2)^{-\Delta_{12}-\Delta_{21}-\lambda_{12}}
\zeta_1^{-\Delta_{13}-\Delta_{31}-\lambda_{13}}
\nonumber \\
\times
\zeta_2^{-\Delta_{23}-\Delta_{32}-\lambda_{23}}
w^{-{{(s_4-2)(s_4-1)}\over2}-\Delta_{41}-\Delta_{42}-\Delta_{43}}
\nonumber \\
\sum^{(s_1-2)\oplus{r};s_2-2;s_3-2;s_4-2}_{ord.{\lbrace{i,j,k}\rbrace};
\lbrace{l,m,n}\rbrace;\lbrace{p,q,r}\rbrace;\lbrace{s,t,u}\rbrace}
\eta_{\mu_{i_1}\nu_{l_1}}...\eta_{\mu_{i_{\lambda_{12}}}\nu_{l_{\lambda_{12}}}}
\nonumber \\
\times
\eta_{\mu_{j_1}\rho_{p_1}}...\eta_{\mu_{i_{\lambda_{13}}}\rho_{p_{\lambda_{13}}}}
\nonumber \\
\eta_{\mu_{k_1}\sigma_{s_1}}...\eta_{\mu_{k_{\lambda_{14}}}\sigma_{s_{\lambda_{14}}}}
\eta_{\nu_{m_1}\rho_{r_1}}...\eta_{{\nu_{m_{\lambda_{23}}}}
\rho_{r_{\lambda_{23}}}}
\nonumber \\
\eta_{\nu_{n_1}\sigma_{t_1}}...\eta_{\nu_{n_{\lambda_{24}}}
\rho_{t_{\lambda_{24}}}}
\eta_{\rho_{r_1}\sigma_{u_1}}...\eta_{\rho_{r_{\lambda_{34}}}}
\sigma_{u_{\lambda_{34}}}
\nonumber \\
(i_1+l_1-2)!...(i_{\lambda_{12}}+l_{\lambda_{12}}-2)!
\nonumber \\
(j_1+p_1-2)!...(j_{\lambda_{13}}+p_{\lambda_{13}}-2)!
\nonumber \\
(k_1+s_1-2)!...(k_{\lambda_{14}}+s_{\lambda_{14}}-2)!
\nonumber \\
(m_1+q_1-2)!...(m_{\lambda_{23}}+q_{\lambda_{23}}-2)!
\nonumber \\
(n_1+t_1-2)!...(n_{\lambda_{24}}+t_{\lambda_{24}}-2)!
\nonumber \\
(r_1+u_1-2)!...(r_{\lambda_{34}}+u_{\lambda_{34}}-2)!
\nonumber \\
\times
(-1)^{s_1+(s_1+\lambda_{12})s_2+s_3(s_4+\lambda_{34})+\lambda_{12}+\lambda_{34}
+\lambda_{13}\lambda_{24}    
+\Delta_{23}+\Delta_{24}+\Delta_{34}}
\nonumber \\
(-1)^{\pi^{(s_1-1);r}(i_1...i_{\lambda_{12}};j_1...j_{\lambda_{13}};k_1...k_{\lambda_{14}})}
\nonumber \\
(-1)^{\pi^{(s_2-2)}({l_1...l_{\lambda_{12}};m_1...m_{\lambda_{23}};n_1...n_{\lambda_{24}}})}
\nonumber \\
\times
(-1)^{\pi^{(s_3-2)}({p_1...p_{\lambda_{13}};q_1...q_{\lambda_{23}};r_1...r_{\lambda_{34}}})}
\nonumber \\
(-1)^{\pi^{(s_4-2)}({s_1...s_{\lambda_{14}};t_1...t_{\lambda_{24}};u_1...u_{\lambda_{34}}})}
\nonumber \\
\equiv
\sum^{(2-\psi)}_{s_1,s_2,s_3,s_4|\lbrace{\lambda_{ij};\Delta_{ij}}\rbrace}
{{\Xi^{(2-\psi)}_{\mu(s_1-2)\nu_(s_2-2)\rho(s_3-2)
\sigma(s_4-2)}}
}
\nonumber \\
\times
{1\over{(\zeta_1-\zeta_2)^{\Delta_{12}+\delta_{21}+\lambda_{12}}
\zeta_1^{\Delta_{13}+\Delta_{31}+\lambda_{13}}
\zeta_2^{\Delta_{23}+\Delta_{32}+\lambda_{23}}
}}
\nonumber \\
\times
w^{-({{(s_4-2)(s_4-1)}\over2}+\Delta_{41}+\Delta_{42}+\Delta_{43})}
\end{align}
where 
$$\Xi^{(2-\psi)}_{\mu(s_1-2|r)\nu_(s_2-2)\rho(s_3-2)\sigma(s_4-2)}
(s_1,s_2,s_3,s_4|\lbrace{\lambda_{ij};\Delta_{ij}}\rbrace)$$
with
$$\mu(s)=\prod_{k=1}^s\mu_k$$
and
$$\mu(s|r)\equiv\mu_1...\mu_{r-1}\mu_{r+1}...\mu_{s};s\geq{r}$$
is again by definition given by the sum  over the orderings defined in the previous
equation, and the summation $\Sigma^{(2-\psi)}$ over the
partitions of $s_i$ in $\lambda$'s  and $\Delta$'s is as explained below.
That is,  the notations are the same as for the pattern 1 and, in addition,
$\sum^{{n\oplus{r}}}_{ord.\lbrace{k}\rbrace}$
denotes  the summation over the possible orderings $(k_1,...k_{n+1}$
of $n+1$ numbers from 1 to $n$ and  $r>n$: (1,2,...,n,r)
and $\pi^{n;r}(k_1,...k_{n+1}$ standing for the number of 
nearest-neighbor pair permutations it takes to make
the ordering $(k_1,...k_{n+1})$ of $n+1$ numbers out of
$(1,2,...,n,r)$.
As  in the pattern 1, the $\lbrace{i,j,k,l,m,n,p,q,r,s,t,u}\rbrace$ elements of the orderings
are taken to satisfy the constraints:
\begin{align}
i_1+...+i_{\lambda_{12}}=\delta_{12}
\nonumber \\
j_1+...+j_{\lambda_{13}}=\delta_{13}
\nonumber \\
k_1+...+k_{\lambda_{14}}=\delta_{14}
\nonumber \\
l_1+...+l_{\lambda_{12}}=\delta_{21}
\nonumber \\
m_1+...+m_{\lambda_{23}}=\delta_{23}
\nonumber \\
n_1+...+j_{\lambda_{24}}=\delta_{24}
\nonumber \\
p_1+...+p_{\lambda_{13}}=\delta_{31}
\nonumber \\
q_1+...+q_{\lambda_{23}}=\delta_{32}
\nonumber \\
r_1+...+r_{\lambda_{34}}=\delta_{34}
\nonumber \\
s_1+...+s_{\lambda_{14}}=\delta_{41}
\nonumber \\
t_1+...+t_{\lambda_{24}}=\delta_{42}
\nonumber \\
u_1+...+u_{\lambda_{34}}=\delta_{43}
\nonumber \\
\sum_{j}\delta_{1j}=(s_1-2)(s_1-3)+r
\nonumber \\
\sum_{j}\delta_{ij}=(s_i-2)(s_i-3)
\end{align}
This concludes the computation of the $\psi$-factor of the $4$-point correlator.
Next, the $X$-factor of (2.2), 
straightforward to compute as well, is again given by the sum over
the following partitions:
\begin{align}
A_X(n_1...n_4|p_1...p_4)
\nonumber \\
\equiv{<}\prod_{i}^{n_1}\partial{X_{\alpha_{i}}}e^{ip_1X}(\zeta_1)
\prod_{j=1}^{n_2}{\partial{X_{\beta_{j}}}e^{ip_2X}}
(\zeta_2)\prod_{k=1}^{n_3}{\partial{X_{\gamma_{k}}}e^{ip_3X}(0)}
\prod_{l=1}^{n_4}{\partial{X_{\delta_{l}}}e^{ip_4X}(w\rightarrow\infty)>}
\nonumber \\
=\sum_{s_1-1,s_2-1,s_3-1,s_4-1|\lbrace\kappa_{IJ}\rbrace;\lbrace{\tau_{IJ}}\rbrace}
{{(s_1-1)!...(s_4-1)!}\over{\prod_{1\leq{I,J}\leq{4};I<J}\kappa_{IJ}!
\tau_{IJ}!\tau_{JI}!}}
\nonumber \\
\times(\zeta_1-\zeta_2)^{-2\kappa_{12}-\tau_{12}-\tau_{21}}
\zeta_1^{-2\kappa_{13}-\tau_{13}-\tau_{31}}
\zeta_2^{-2\kappa_{23}-\tau_{23}-\tau_{32}}
w^{-s_4+1-\kappa_{41}-\kappa_{42}-\kappa{43}-\tau_{41}-\tau_{42}-\tau_{43}}
\nonumber \\
\times\prod_{{\lbrace}1\leq{k_1}\leq{\kappa_{12}}\rbrace}
\prod_{\lbrace\kappa_{12}+1\leq{k_2}\leq{\kappa_{12}+\kappa_{13}}\rbrace}
\prod_{\lbrace\kappa_{12}+\kappa_{13}+1\leq{k_3}\leq{\kappa_{12}+\kappa_{13}+\kappa_{14}}\rbrace}
\nonumber \\
\prod_{\lbrace\kappa_{12}+\kappa_{13}+\kappa_{14}+1\leq{k_4}\leq\kappa_{12}
+\kappa_{13}+\kappa_{14}+\tau_{12}\rbrace}
\prod_{\lbrace\kappa_{12}+\kappa_{13}+
\kappa_{14}+\tau_{12}+1\leq{k_5}\leq\kappa_{12}+\kappa_{13}+\kappa_{14}+\tau_{12}+\tau_{13}\rbrace}
\nonumber \\
\prod_{\lbrace\kappa_{12}+\kappa_{13}+\kappa_{14}
+\tau_{12}+{\tau_{13}+1\leq{k_6}\leq{s_1-1}}\rbrace}
\prod_{{\lbrace}1\leq{l_1}\leq{\kappa_{12}}\rbrace}
\prod_{\lbrace\kappa_{12}+1\leq{l_2}\leq{\kappa_{12}+\kappa_{23}}\rbrace}
\prod_{\lbrace\kappa_{12}+\kappa_{23}+1\leq{l_3}\leq{\kappa_{12}+\kappa_{23}+\kappa_{24}}\rbrace}
\nonumber \\
\prod_{\lbrace\kappa_{12}+\kappa_{23}+\kappa_{24}+1\leq{l_4}\leq\kappa_{12}
+\kappa_{23}+\kappa_{24}+\tau_{21}\rbrace}
\prod_{\lbrace\kappa_{12}+\kappa_{23}+   
\kappa_{24}+\tau_{21}+1\leq{l_5}\leq\kappa_{12}+\kappa_{23}
+\kappa_{24}+\tau_{21}+\tau_{23}\rbrace}
\nonumber \\
\prod_{\lbrace\kappa_{12}+\kappa_{23}+\kappa_{24}
+\tau_{21}+{\tau_{23}+1\leq{l_6}\leq{s_2-1}}\rbrace}
\prod_{{\lbrace}1\leq{m_1}\leq{\kappa_{13}}\rbrace}
\prod_{\lbrace\kappa_{13}+1\leq{m_2}\leq{\kappa_{13}+\kappa_{23}}\rbrace}
\nonumber \\
\prod_{\kappa_{13}+\kappa_{23}+1\leq{m_3}\leq{\kappa_{13}+\kappa_{23}+\kappa_{34}}}
\prod_{\kappa_{13}+\kappa_{23}+\kappa_{34}+1\leq{m_4}\leq
\kappa_{13}+\kappa_{23}+\kappa_{34}+\tau_{31}}
\nonumber \\
\prod_{\lbrace\kappa_{13}
+\kappa_{23}+                  
\kappa_{34}+\tau_{31}+1\leq{m_5}\leq\kappa_{13}+
\kappa_{23}                                                              
+\kappa_{34}+\tau_{31}+\tau_{32}\rbrace}
\prod_{\lbrace\kappa_{12}+\kappa_{23}                                            
+\kappa_{34}                                                                    
+\tau_{31}+{\tau_{32}+1\leq{m_6}
\leq{s_3-1}}\rbrace}
\nonumber \\
\prod_{{\lbrace}1\leq{n_1}\leq{\kappa_{14}}\rbrace}
\prod_{\kappa_{14}+1\leq{n_2}\leq{\kappa_{14}+\kappa_{24}}}
\prod_{\lbrace\kappa_{14}+\kappa_{24}+1\leq{n_3}\leq{\kappa_{14}+\kappa_{24}+\kappa_{34}}\rbrace}
\nonumber \\
\prod_{\lbrace\kappa_{14}+\kappa_{24}+\kappa_{34}+1\leq{n_4}\leq                                       
\kappa_{14}+\kappa_{24}+\kappa_{34}+\tau_{41}\rbrace}
\prod_{\lbrace\kappa_{14}
+\kappa_{24}+
\kappa_{34}+\tau_{41}+1\leq{n_5}\leq\kappa_{14}+
\kappa_{24}
+\kappa_{34}+\tau_{41}+\tau_{42}\rbrace}
\nonumber \\
\prod_{\lbrace\kappa_{14}+\kappa_{24}  
+\kappa_{34}
+\tau_{41}+{\tau_{42}+1\leq{n_6}
\leq{s_4-1}}\rbrace}
\nonumber \\
\times\eta_{\alpha_{k_1\beta_{l_1}}}\eta_{\alpha_{k_2\gamma_{m_1}}}
\eta_{\alpha_{k_3}\delta_{n_1}}\eta_{\beta_{l_2}\gamma_{m_2}}\eta_{\beta_{l_3}\delta_{n_2}}
\nonumber \\
\eta_{\gamma_{m_3}\delta_{n_3}}(ip_2)_{\alpha_{k_4}}(ip_3)_{\alpha_{k_5}}(ip_4)_{\alpha_{k_6}}
(-ip_1)_{\beta_{l_4}}(ip_3)_{\beta_{l_5}}(ip_4)_{\beta_{l_6}}(-ip_1)_{\gamma_{m_4}}(-ip_2)_{\gamma_{m_5}}
\nonumber \\
(ip_4)_{\gamma_{m_6}}(-ip_1)_{\delta_{n_4}}(-ip_2)_{\delta_{n_5}}(-ip_3)_{\delta_{n_6}}
\times\delta(p_1+p_2+p_3+p_4)
\nonumber 
\end{align}
\begin{align}
\equiv
\sum_{s_1-1,s_2-1,s_3-1,s_4-1|\lbrace\kappa_{IJ}\rbrace;\lbrace{\tau_{IJ}}\rbrace}
{{(s_1-1)!...(s_4-1)!}\over{\prod_{1\leq{I,J}\leq{4};I<J}\kappa_{IJ}!
\tau_{IJ}!\tau_{JI}!}}
\nonumber \\
(\zeta_1-\zeta_2)^{-2\kappa_{12}-\tau_{12}-\tau_{21}}
\zeta_1^{-2\kappa_{13}-\tau_{13}-\tau_{31}}
\nonumber \\
\times\zeta_2^{-2\kappa_{23}-\tau_{23}-\tau_{32}}
w^{-s_4+1-\kappa_{41}-\kappa_{42}-\kappa{43}-\tau_{41}-\tau_{42}-\tau_{43}}
\nonumber \\
\times\Xi^{(X)}_{\alpha(s_1-1)\beta(s_2-1)\gamma(s_3-1)\delta(s_4-1)}
(s_1-1,...s_4-1|\lbrace\kappa\rbrace;\lbrace\tau\rbrace)
\end{align}
where the factors
 $\Xi^{(X)}(n_1...n_4|\lbrace\kappa\rbrace;\lbrace\tau\rbrace)$ are by definition
introduced according to (4.126) 

and the sum 
$\sum_{n_1,n_2,n_3,n_4|\lbrace\kappa_{IJ}\rbrace;\lbrace{\tau_{IJ}}\rbrace}$ is
 taken over the non-ordered partitions:
\begin{align}
n_I=\sum_{1\leq{J}\leq{4};J{\neq}I}(\kappa_{IJ}+\tau_{IJ})
\nonumber \\
\kappa_{IJ}=\kappa_{JI}
\end{align}
The final ingredient for the higher-spin amplitude comes from the ghost factor.
First of all, note that, since the Bell polynomials in the ghost fields appearing in
 the expressions
for the vertex operators are limited to $B^{(m)}_{\phi-\chi}$ and $B^{(n)}_{2\phi-2\chi-\sigma}$
and the operator products between the derivatives of $\phi-\chi$ and $2\phi-2\chi-\sigma$
are nonsingular, the polynomials only contract with the ghost exponentials.
The pattern for the ghost part of the correlator is
given by
\begin{align}
A_{gh}(s_1...s_4|n)
\nonumber \\
=<:ce^{\chi+(s_1-3)\phi}B^{(n)}_{\phi-\chi}:(\zeta_1)
:e^{(s_2-2)\phi}B^{(2s_2-4)}_{2\phi-2\chi-\sigma}:(\zeta_2)
\nonumber \\
ce^{-s_3\phi}(0)ce^{-s_4\phi}(w\rightarrow\infty)>
\nonumber \\
=\sum_{{\lbrace}n|\omega_{12},\omega_{13},\omega_{14}\rbrace}
\sum_{{\lbrace}2s_2-4|\omega_{21},\omega_{23},\omega_{24}\rbrace}
\Xi^{(gh)}(s_1...s_4|n;\lbrace{\omega}\rbrace)
\nonumber \\
{\times}(\zeta_1-\zeta_2)^{-(s_1-3)(s_2-2)-\omega_{12}-\omega_{21}}
\zeta_1^{(s_1-3)s_3+1-\omega_{13}}
\nonumber \\
\times\zeta_2^{(s_2-2)s_3+1-\omega_{23}}
w^{(s_1+s_2-5)s_4+2-\omega_{14}-\omega_{24}}
\end{align}
where
\begin{align}
\Xi^{(gh)}(s_1...s_4|n;\lbrace{\omega}\rbrace)
=C(2-s_2|n;n-\omega_{12})
\nonumber \\
\times
C(s_3|n-\omega_{12};n-\omega_{12}-\omega_{13})
\nonumber \\
\times
C(s_4|n-\omega_{12};n-\omega_{12}-\omega_{13};0)
\nonumber \\
\times
C(3-2s_1|2s_2-4;2s_2-4-\omega_{21})
\nonumber \\
\times
C(2s_3-1
|2s_2-4-\omega_{21};2s_2-4-\omega_{21}-\omega_{23})
\nonumber \\
{\times}C(2s_4-1|2s_2-4-\omega_{12}-\omega_{23};0)
\end{align}
where the coefficients
$C(q|n;n-m)$ stem from the operator product:
\begin{align}
B^{(N)}_{\alpha_1\phi+\alpha_2\chi+\alpha_3\sigma}(z)e^{\beta_1\phi+\beta_2\chi+\beta_3\sigma}(w)
 \nonumber \\
=\sum_{n=0}^N(z-w)^{-n}{C(-\alpha_1\beta_1+\alpha_2\beta_2+\alpha_3\beta_3|N;N-n)}
\nonumber \\
:B^{(N-n)}_{\alpha_1\phi+\alpha_2\chi+\alpha_3\sigma}(z)e^{\beta_1\phi+\beta_2\chi+\beta_3\sigma}:(w)
\end{align}
with
\begin{align}
C(a|N;N-n)={{\Gamma(1+a)}\over{n!\Gamma(1+a-n)}}
\end{align}
and $\omega_{ij}$-numbers satisfy the constraints
\begin{align}
\omega_{12}+\omega_{13}+\omega_{14}=n
\nonumber \\
\omega_{21}+\omega_{23}+\omega_{24}=2s_2-4
\end{align}
and define the non-ordered partitions of $n$ and $2s_2-4$,
with the summations taken over these partitions.
This concludes the computation of all of the patterns ($\psi$,$X$, and ghost) contributing
to the $4$-point correlation function of the higher spins.
Finally, using 
\begin{align}
\int_0^1{dz_1}\int_{0}^{z_1}dz_2{z_1}^az_2^b(z_1-z_2)^c={{\Gamma(a+1)\Gamma(c+1)}\over
{(a+b+c+2)\Gamma(a+c+2)}}
\end{align},
substituting the patterns into the integrals in (4.116) and integrating, we obtain the 
following
answer for the amplitude:
\begin{align}
A(s_1,s_2,s_3,s_4)=
{{\Omega^{s_1|s_1-3}(p_1)\Omega^{s_2|s_2-3}(p_2)\Omega^{s_3|s_3-3}(p_3)\Omega^{s_4|s_4-3}(p_4)}\over{8}}
\nonumber \\
\times
\lbrace\sum_{q=1}^{s_1-2}(-1)^q(q-1)!
\sum_{{\lbrace}2s_2-4|\omega_{21},\omega_{23},\omega_{24}\rbrace}
\sum_{{\lbrace}s_1+q+2|\omega_{12},\omega_{13},\omega_{14}\rbrace}
\sum_{{\lbrace}s_1-3,s_2-2,s_3-2,s_4-2|\lbrace\lambda_{ij}\rbrace\rbrace}
\nonumber \\
\sum_{{{(s_1-2)(s_1-3)}\over2}-q;
{{(s_2-2)(s_2-3)}\over2};{{(s_3-2)(s_2-3)}\over2};
{{(s_4-2)(s_4-3)}\over2}|\lbrace|\Delta_{ij}\rbrace}
\nonumber \\
(\sum_{s_1-2;s_2-1;s_3-1;s_4-1|\lbrace\tau\rbrace;\lbrace\kappa\rbrace}
{{(s_1-1)!...(s_4-1)!}\over{\prod_{1\leq{I,J}\leq{4};I<J}\kappa_{IJ}!
\tau_{IJ}!\tau_{JI}!}}
\nonumber \\
\Xi^{(1-\psi)}_{\mu(s_1-2;q)\nu(s_2-2)\rho(s_3-2)   
\sigma(s_4-2)}(\lbrace{\lambda\rbrace,\lbrace{\Delta}\rbrace})
\nonumber \\
(-\eta_{\mu_q\alpha_q}
\Xi^{(X)}_{\alpha(s_1-1|q)\beta(s_2-1)\gamma(s_3-1)\delta(s_4-1)}
(s_1-1,...s_4-1|\lbrace\kappa\rbrace;\lbrace\tau\rbrace)
\nonumber \\
-\sum_{s_1-1;s_2-1;s_3-1;s_4-1|\lbrace\tau\rbrace;\lbrace\kappa\rbrace}
\nonumber \\
(ip_1)_{\mu_q}
\Xi^{(X)}_{\alpha(s_1-1)\beta(s_2-1)\gamma(s_3-1)\delta(s_4-1)}
(s_1-1,...s_4-1|\lbrace\kappa\rbrace;\lbrace\tau\rbrace)
\nonumber \\
\Xi^{(gh)}(s_1...s_4|s_1+q+2;\lbrace{\omega}\rbrace)
F(p_1...p_4|\lbrace\lambda,\Delta,\kappa,\tau,\omega\rbrace))
\nonumber \\
+
\sum_{q=1}^{s_1-2}\sum_{r=0}^{s_1+q}(-1)^q(q-1)!
\sum_{{\lbrace}2s_2-4|\omega_{21},\omega_{23},\omega_{24}\rbrace}
\sum_{{\lbrace}s_1+q-r|\omega_{12},\omega_{13},\omega_{14}\rbrace}
\nonumber \\
\sum_{{\lbrace}s_1-3,s_2-2,s_3-2,s_4-2|\lbrace\lambda_{ij}\rbrace\rbrace}
\sum_{{\lbrace}{{(s_1-2)(s_1-3)}\over2}-q;      
{{(s_2-2)(s_2-3)}\over2};
{{(s_3-2)(s_2-3)}\over2};                                                                          
{{(s_4-2)(s_4-3)}\over2}|\lbrace|\Delta_{ij}\rbrace\rbrace}
\nonumber \\
{\times}(\sum_{s_1-2;s_2-1;s_3-1;s_4-1|\lbrace\tau\rbrace;\lbrace\kappa\rbrace}
(-1)^q(q-1)!
\nonumber \\
{{(s_1-1)!...(s_4-1)!}\over{\prod_{1\leq{I,J}\leq{4};I<J}\kappa_{IJ}!
\tau_{IJ}!\tau_{JI}!}}
\Xi^{(1-\psi)}_{\mu(s_1-2|q)\nu(s_2-2)\rho(s_3-2)\sigma(s_4-2)}
(\lbrace{\lambda\rbrace,\lbrace{\Delta}\rbrace})
\nonumber \\
\times
\Xi^{(gh)}(s_1...s_4|s_1+q-r;\lbrace{\omega}\rbrace)
\nonumber \\ 
\times\lbrack
((-1)^{r+1}\Xi^{(X)}_{\alpha(s_1-1)\beta(s_2-1|q)\gamma(s_3-1)\delta(s_4-1)}
(s_1-1,...s_4-1|\lbrace\kappa\rbrace;\lbrace\tau\rbrace)
\nonumber \\
\times
(\eta_{\alpha_q\beta_q}(r+1)!F_r^{(1)}(p_1...p_4|\lbrace\lambda,\Delta,\kappa,\tau,\omega\rbrace)
\nonumber \\
+(-1)^{r+1}\Xi^{(X)}_{\alpha(s_1-1)\beta(s_2-1)\gamma(s_3)\delta(s_4-1)}
(s_1-1,...s_4-1|\lbrace\kappa\rbrace;\lbrace\tau\rbrace)
(ip_2)^qr!
\nonumber \\
\times
F_{r-1}^{(1)}(p_1...p_4|\lbrace\lambda,\Delta,\kappa,\tau,\omega\rbrace)
(s_1-1,...s_4-1|\lbrace\kappa\rbrace;\lbrace\tau\rbrace)))
\nonumber \\
+
((-1)^{r+1}\Xi^{(X)}_{\alpha(s_1-1)\beta(s_2-1)\gamma(s_3-1|q)\delta(s_4-1)}
(s_1-1,...s_4-1|\lbrace\kappa\rbrace;\lbrace\tau\rbrace)
\nonumber \\
\times
(\eta_{\alpha_q\gamma_q}(r+1)!F_r^{(2)}(p_1...p_4|\lbrace\lambda,\Delta,\kappa,\tau,\omega\rbrace)
\nonumber \\
+(-1)^{r+1}\Xi^{(X)}_{\alpha(s_1-1)\beta(s_2-1)\gamma(s_3)\delta(s_4-1)}
(s_1-1,...s_4-1|\lbrace\kappa\rbrace;\lbrace\tau\rbrace)
\nonumber \\
\times
(ip_3)^qr!F_{r-1}^{(2)}(p_1...p_4|\lbrace\lambda,\Delta,\kappa,\tau,\omega\rbrace)
(s_1-1,...s_4-1|\lbrace\kappa\rbrace;\lbrace\tau\rbrace)))
\nonumber 
\end{align}
\begin{align}
+
((-1)^{r+1}\Xi^{(X)}_{\alpha(s_1-1)\beta(s_2-1)\gamma(s_3-1)\delta(s_4-1|q)}
\nonumber \\
\times
(\eta_{\alpha_q\delta_q}(r+1)!F_r^{(3)}(p_1...p_4|\lbrace\lambda,\Delta,\kappa,\tau,\omega\rbrace)
\nonumber \\
+(-1)^{r+1}\Xi^{(X)}_{\alpha(s_1-1)\beta(s_2-1)\gamma(s_3)\delta(s_4-1)}
(s_1-1,...s_4-1|\lbrace\kappa\rbrace;\lbrace\tau\rbrace)
\nonumber \\
\times
(ip_4)^qr!F_{r-1}^{(3)}(p_1...p_4|\lbrace\lambda,\Delta,\kappa,\tau,\omega\rbrace)
))
\rbrack
\nonumber \\
-
\sum_{r={s_1-1}}^{s_1+2}
\sum_{{\lbrace}s_1+2-r|\omega_{12},\omega_{13},\omega_{14}\rbrace}
\nonumber \\
\sum_{{\lbrace}2s_2-4|\omega_{21},\omega_{23},\omega_{24}\rbrace}
\sum_{{\lbrace}s_1+q+2|\omega_{12},\omega_{13},\omega_{14}\rbrace}
\nonumber \\
\sum_{{\lbrace}s_1-1,s_2-2,s_3-2,s_4-2|\lbrace\lambda_{ij}\rbrace\rbrace}
\nonumber \\
\sum_{{\lbrace}{{(s_1-2)(s_1-3)}\over2}+r;                                                              
{{(s_2-2)(s_2-3)}\over2};{{(s_3-2)(s_2-3)}\over2};                                            
{{(s_4-2)(s_4-3)}\over2}|\lbrace|\Delta_{ij}\rbrace\rbrace}
\nonumber \\
{1\over{r!}}
{{(s_1-1)!...(s_4-1)!}\over{\prod_{1\leq{I,J}\leq{4};I<J}\kappa_{IJ}!
\tau_{IJ}!\tau_{JI}!}}
\nonumber \\
{\times}{\Xi^{(2-\psi)}_{\mu_r|\mu(s_1-2)\nu_(s_2-2)\rho(s_3-2)                                        
\sigma(s_4-2)}(s_1,s_2,s_3,s_4|\lbrace{\lambda_{ij};\Delta_{ij}}\rbrace)}
\nonumber \\
\times\Xi^{(X)}_{\alpha(s_1-1|r)\beta(s_2-1)\gamma(s_3-1)\delta(s_4-1)}
(s_1-1,...s_4-1|\lbrace\kappa\rbrace;\lbrace\tau\rbrace)
\nonumber \\
\times
\Xi^{(gh)}(s_1...s_4|s_1+2-r;\lbrace{\omega}\rbrace)
F(p_1...p_4|\lbrace\lambda,\Delta,\kappa,\tau,\omega\rbrace)
\nonumber \\
-
\sum_{r={s_1-1}}^{s_1+1}
\sum_{{\lbrace}n|\omega_{12},\omega_{13},\omega_{14}\rbrace}
\sum_{{\lbrace}2s_2-4|\omega_{21},\omega_{23},\omega_{24}\rbrace}
\nonumber \\
\sum_{{\lbrace}s_1+1-r|\omega_{12},\omega_{13},\omega_{14}\rbrace}
\sum_{{\lbrace}s_1-3,s_2-2,s_3-2,s_4-2|\lbrace\lambda_{ij}\rbrace\rbrace}
\nonumber \\
\sum_{\lbrace{{(s_1-2)(s_1-3)}\over2}+r;                                                                  
{{(s_2-2)(s_2-3)}\over2};{{(s_3-2)(s_2-3)}\over2};                                              
{{(s_4-2)(s_4-3)}\over2}|\lbrace|\Delta_{ij}\rbrace\rbrace}
\nonumber \\
{1\over{r!}}
{{(s_1-1)!...(s_4-1)!}\over{\prod_{1\leq{I,J}\leq{4};I<J}\kappa_{IJ}!
\tau_{IJ}!\tau_{JI}!}}
(ip^{\mu_r})
\nonumber \\
\times
{\Xi^{(2-\psi)}_{\mu_r;\mu(s_1-2)\nu_(s_2-2)\rho(s_3-2)
\sigma(s_4-2)}(s_1,s_2,s_3,s_4|\lbrace{\lambda_{ij};\Delta_{ij}}\rbrace)}
\nonumber \\
\times\Xi^{(X)}_{\alpha(s_1-1)\beta(s_2-1)\gamma(s_3-1)\delta(s_4-1)}
(s_1-1,...s_4-1|\lbrace\kappa\rbrace;\lbrace\tau\rbrace)
\nonumber \\
\times
\Xi^{(gh)}(s_1...s_4|s_1+1-r;\lbrace{\omega}\rbrace)
F(p_1...p_4|\lbrace\lambda,\Delta,\kappa,\tau,\omega\rbrace)
\end{align}

where

\begin{align}
F(p_1...p_4|\lbrace\lambda,\Delta,\kappa,\tau,\omega\rbrace)
\equiv
F(A;B;C)
\nonumber \\
={{\Gamma(1+C(\lbrace\lambda,\Delta,\tau,\gamma,\omega,\lbrace{s}\rbrace\rbrace))
\Gamma(1+A(\lbrace\lambda,\Delta,\tau,\gamma,\omega,\lbrace{s}\rbrace\rbrace))}\over
{(A+B+C+2)\Gamma(A+C+2)}}
\end{align}

with
\begin{align}
C=
p_1p_2-\Delta_{12}-\Delta_{21}-\lambda_{12}-\tau_{12}-\tau_{21}
\nonumber \\
-2\kappa_{12}-(s_1-3)(s_2-2)-\omega_{12}-\omega_{21}
\nonumber \\
A=
p_1p_3-\Delta_{13}-\Delta_{31}-\lambda_{13}-\tau_{13}-\tau_{31}
\nonumber \\
-2\kappa_{13}+(s_1-3)s_3-\omega_{13}
\nonumber \\
B=
-{1\over2}(p_1p_4+p_2^2+p_3^2+p_4^2)
-\omega_{23}
\nonumber \\
-\Delta_{23}-\Delta_{32}-\lambda_{23}-\tau_{23}-\tau_{32}-2\kappa_{23}
+(s_2-2)s_3+1)
\end{align}

and 
\begin{align}
F_r^{(1)}(A;B;C)=F(A;B;C+r)\nonumber \\
F_r^{(2)}(A;B;C)=F(A+r;B;C) \nonumber \\
F_r^{(3)}(A;B;C)=F(A;B+r;C)
\end{align}
Finally, all the summations over the partitions in the
higher-spin  amplitude
 are subject to one more unitarity constraint:
\begin{align}
s_4(s_1+s_2-6)+3-{1\over2}(s_4-2)(s_4-1)
\nonumber \\
-\sum_{j=1}^3(\Delta_{4j}
-\kappa_{4j}-\tau_{4j})-\omega_{14}-\omega_{24}=0
\end{align}
which stems from the condition that, in the limit
 $w\rightarrow\infty$, as the location of the $s_4$ vertex operator
taken to the infinity, only the terms behaving as ${\sim}w^0$ survive.

This concludes the derivation of the general 4-point amplitude for the 
higher spins in AdS, provided that  the spin values satisfy the constraint (4.115)
and the particles are polarized and propagating along the $AdS$ boundary.
In the next section we shall discuss the relevance of this  amplitude to the higher-spin
quartic interactions in AdS (restricted to rectangluar case and propagation along the $AdS$ boundary.  

\section{\bf   Flows from the Cubics : No-Go Constraints }

The field theory limit ($p_ip_j\rightarrow{0}$) of  the amplitude (2.20)  does not by itself
automatically reproduce the full  quartic interaction in the low-energy effective 
action: it is only invariant under the linearized  gauge transformations, induced by BRST operator
in the free theory. This operator is deformed by the worldsheet RG flow in the leading order, and therefore
the full gauge symmetry of the higher-spin action at the quartic order is realized nonlinearly. To obtain the quartic interaction
terms invariant under the full gauge symmetry, one has to carefully combine the field-theoretic limit of the amplitude (2.20)
with contributions stemming from the deformations of the cubic terms under the worldsheet RG flow, leading to quartic counterterms.
This makes a problem of extracting higher-spin quartic interactions from string theory quite cumbersome, since
the number of such contriburtions is generally infinite (as the higher-spin algebra is infinite). In general, this
infinite number of counterterms is  one possible source of nonlocalities in quartic higher-spin interactions; another,as has been noted before,
is related to the ghost coupling  structure of the higher-spin vertex operators. In the rectangular limit, however, things get simplified due
to ghost cohomology conditions combined with the ghost number selection rules, which
impose stringent constraints on the correlators. As  a result of those constraints,
all the 3-point higher-spin correlation functions, relevant to flow contributions to the rectangular quartic terms, 
turn out to vanish. As a result, the rectangular limit of the quartic interactions is determined solely by the corresponding
4-point correlators and the nonlocality structure of such interactions is entirely encrypted in the ghost structure 
of the higher-spin vertex operators.
To see this, consider a rectangular $4$-point  correlator 
$$A(s_1,...s_4)=<W_{s_1}^{(+)}W_{s_2}^{(+)}W_{s_3}^{(-)}W_{s_4}^{(-)}>\Phi_{s_1}...\Phi_{s_4}$$
subject to the rectangular constraint $s_1+s_2=s_3+s_4+3$

(as before, we adopt the convention $V_s=\Phi_sW_s$ where $\Phi$ are the space-time fields and $W$ are the worldsheet
operators)
There are several possibilities for the quartic terms  to emerge as RG flows of the cubics in the low-energy action.
We will consider one of them (others can be treated similarly).
One such type of terms is $\sim\Phi_{s_1}\Phi_{s_2}\Phi_\sigma$ where $\sigma$ is some spin value.
The $b-c$ and $\phi$-ghost number balance requires that terms are generated by the correlation functions
\begin{eqnarray}
<V_{s_1}^{(+)}V_{s_2}^{(-)}V_{\sigma}^{(-)}>=
<V_{s_1}^{(s_1-3)}V_{s_2}^{(-s_2)}V_\sigma^{(s_2-s_1+1)}>
\nonumber \\
=
<V_{s_1}^{(s_1-3)}V_{s_2}^{(-s_2)}:\Gamma^{\sigma+1+s_2-s_1}V_\sigma^{(-\sigma)}:>
\end{eqnarray}
with numbers in superscripts referring to the $\phi$-ghost numbers of the operators.
Now, since $<V_{s_1}^{(s_1-3)}\in{H_{s_1-2}}\sim{H_{-s_1}}$ and
$V_{s_2}^{(-s_2)}\in{H_{-s_2}}\sim{H_{s_2-2}}$, the first operator is annihilated by
$\Gamma^{-1}$ and the second by $\Gamma$. This immediately entails the non-vanishing
constraint for the correlator:
\begin{eqnarray}
\sigma=s_1-s_2-1
\end{eqnarray}
since otherwise either inverse or direct picture-changing operators can be moved
from the location of $V_\sigma$ to the locations of $V_{s_1}$ or $V_{s_2}$
respectively, thus annihilating the correlator. In other words, the non-vanishing constraint
requires that $V_\sigma\in{H_{-\sigma}}$ must be taken at its minimal negative picture $-\sigma$.
Now consider the RG flow of the cubic $\Phi_{s_1}\Phi_{s_2}\Phi_\sigma$ into the quartic
$\Phi_{s_1}\Phi_{s_2}\Phi_{s_3}\Phi_{s_4}$. This flow clearly stems from the quadratic contribution
to the worldsheet $\beta$-function of $\Phi_\sigma$ given by
\begin{eqnarray}
\beta_{\Phi_\sigma}\sim\alpha\Phi_{s_3}\Phi_{s_4}
\end{eqnarray}
where $\alpha$ is the structure constant given by the 3-point correlator
$<W_{s_3}W_{s_4}W_\sigma>$.
Since $\Phi_{s_3}$ and $\Phi_{s_4}$ are the frame-like higher-spin extra fields with the 
structure $\Phi_{s}\sim\Omega_{s|s-3}$ in the low-energy action, their vertex operators can be taken
at their minimal pictures $s_3-2$ and $-s_4$. 
Furthermore, $V_\sigma$-operator can be taken to the canonical picture $-\sigma$,
using $|s_4-s_3+1+\sigma|$ generalized zero torsion constraints on $\Phi_\sigma$
denoted, for simplicity, as $\Phi_\sigma\rightarrow{\tilde\Phi}_\sigma$
The $b-c$ and $\phi$-ghost number constraints then require
\begin{eqnarray}
\alpha=<W_{s_3}^{(+)}W_{s_4}^{(-)}W_{\sigma}^{(-)}>\Phi_{s_1}\Phi_{s_2}\Phi_\sigma
=
<W_{s_3}^{(s_3-3)}W_{s_4}^{(-s_4)}W_{\sigma}^{(s_4-s_3+1)}>\Phi_{s_1}\Phi_{s_2}\Phi_\sigma
\nonumber \\
=<W_{s_3}^{(s_3-3)}W_{s_4}^{(-s_4)}W_{\sigma}^{-\sigma}>\Phi_{s_1}\Phi_{s_2}{\tilde\Phi}_\sigma
\end{eqnarray}
which entails another non-vanishing constraint:
\begin{eqnarray}
\sigma=s_3-s_4-1
\end{eqnarray}
Comparing two non-vanishing constraints above leads,
along with the rectangular constraints to
\begin{eqnarray}
s_1-s_2=s_3-s_4
\nonumber \\
s_1+s_2=s_3+s_4+3
\end{eqnarray}
implying, in particular,
\begin{equation}
s_1=s_3+{3\over2}
\end{equation}
but this contradicts our initial assumption that all the operators are
in the NS sector, i.e. all the spins are integer.
This  concludes the proof of no-flow property of the rectangular $4$-vertices.
In particular, this property ensures that the $4$-point amplitude (2.20) 
describes
the quartic terms invariant under the
full (nonlinear) gauge symmetry transformations for the higher-spin fields.

\section{\bf Conclusion and Discussion}

In this work we have analysed a sufficiently large class of quartic  higher-spin interactions
for arbitrary spin values an $AdS_5$, only subject to the rectangular constraint and propagating along the AdS

We found that the nonlocality structure of the  interactions is closely related to the ghost structure
of the vertex operators describing the propagation of the massless higher-spin modes along the $AdS_5$ boundary,
and this structure is universal for all the 4-vertices satisfying the rectangular constraint.
The remarkable simplification in the structure of the rectangular 4-vertices is  that, due to
the ghost cohomology conditions combined with the rectangular limit, they receive no
contributions from the RG flows of the cubic vertices, making it possible to deduce the 
quartic interactions directly from the 4-point amplitudes in ``larger'' string theory.
It would be interesting to check if this no-flow structure persists in some other dimensions, where the manifest
expressions for the higher-spin vertex operators are  more complicated, as well to generalize things to the case of
arbitrary propagation in the bulk. Holographically,  the quartic higher-spin interactions in AdS space
must be related  to conformal blocks in the boundary CFT. It is not a trivial relation,
given the apparent nonlocality of the higher-spin interactions versus local interactions in the boundary CFT.
We hope that the rectangular case, which strongly simplifies the structure of the $4$-vertices for higher spin,
may be a useful toy model to approach this question. Can we  interpret the 
no-flow phenomenon in terms of the boundary CFT? 
Is it related, in some way, to some special class  of solutions of Vasiliev's equations, not yet found?
We hope to address these, and many other issues in the works to come.

\section{\bf Acknowledgements}

The authors gratefully acknowledge the support of National Science Foundation of China (NSFC) under the project
11575119.

\end{document}